\documentclass[%
 aip,
 amsmath,amssymb,
 reprint,%
]{revtex4-1}

\usepackage{graphicx}% Include figure files
\usepackage{dcolumn}% Align table columns on decimal point
\usepackage{bm}% bold math
\usepackage[utf8]{inputenc}
\usepackage[T1]{fontenc}
\usepackage{mathptmx}
\usepackage{etoolbox}
\usepackage{multirow}

\makeatletter
\def\@email#1#2{%
 \endgroup
 \patchcmd{\titleblock@produce}
  {\frontmatter@RRAPformat}
  {\frontmatter@RRAPformat{\produce@RRAP{*#1\href{mailto:#2}{#2}}}\frontmatter@RRAPformat}
  {}{}
}%
\makeatother
\begin{document}

\preprint{AIP/123-QED}

\title[Investigation of Rb$^+$ Milling Rates using an Ultracold Focused Ion Beam]{Investigation of Rb$^+$ Milling Rates using an Ultracold Focused Ion Beam}
% Force line breaks with \\
\author{S. Xu, Y. Li, and E. J. D. Vredenbregt*}
    \email{e.j.d.vredenbregt@tue.nl}
\affiliation{ Department of Applied Physics, Eindhoven University of Technology, P.O. Box 513, 5600 MB Eindhoven, the Netherlands}%
\date{\today}% It is always \today, today,

\begin{abstract}
Several ion source alternatives for current focused ion beam (FIB) systems have been studied to achieve higher brightness, including cold atom ion sources. However, a study of ultracold ions interacting with often used materials is seldom reported. Here we investigate milling on several typical samples in a prototype ultracold Rb FIB system at 8.5 keV beam energy. For polycrystalline metallic substrates, such as Cu and Au, patterns milled by Rb$^+$ ions are observed to have reduced surface roughness, but still high milling rates compared with those milled by Ga$^+$ ions. Rb$^+$ also shows similar sputter rates as 30 keV Ga$^+$ on semiconductor substrates GaAs and InP. Special cases for Rb$^+$ milling show that the Rb$^+$ ion beam has a $2.6 \times$ faster sputter rate on diamond but a $3 \times$ slower sputter rate on Al compared with a normal 30 keV Ga$^+$ ion beam. Generally, a Rb$^+$ ion beam is shown to be suitable for nanostructuring of several basic materials.
\end{abstract}

\maketitle

\section{\label{sec:intro}Introduction}

Over the past 20 years, FIBs have been widely used for nanoscale manufacturing and characterization techniques in the materials science and semiconductor industry. FIBs serve as a reliable tool mostly in failure analysis\cite{nikawa1991applications}, transmission electron microscopy (TEM) sample preparation\cite{giannuzzi1999review} and integrated circuit (IC) edit\cite{harriott1986integrated, melngailis1986focused}, and the three applications conversely drive FIB tool development simultaneously. In the semiconductor industry, the nanomachining precision of FIB is required to keep pace with decreased feature sizes by Moore's law (transistor density doubles per 2 years)\cite{moore1965moore}. Currently, the Ga liquid metal ion source (LMIS), as a representative of commercial FIBs, is nearly running out of milling resolution, while IBM has already reported a 2 nm process. Thus, a new alternative ion source FIB with a high sputter yield and smaller beam size is required.
\par
Recently, several ion source alternatives for Ga LMIS have been commercialized. The gas field ionization source (GFIS)\cite{tondare2005quest} is the most promising one, which mainly operates with light ion species, such as He\cite{ward2006helium, morgan2006introduction} and Ne\cite{livengood2011neon}. Such an ion source has higher brightness than the Ga LMIS, thus producing a smaller spot size and higher image resolution correspondingly\cite{hill2011advances}. Researches have shown helium ion microscopy has a low sputter yield and He$^+$ irradiation can cause large-area subsurface damage of He nano-bubbles in a silicon substrate\cite{livengood2009subsurface, li2019study}, which limits the applications in nanomachining. Another alternative is the inductively coupled plasma FIB, with the Xe FIB as representative. It shows quick sputter rates and can be applied for large volume solid machining by $\mu$A beam current\cite{burnett2016large}. TEM characterization also confirms Xe$^+$ can reduce subsurface amorphous thickness\cite{kelley2013xe+}. But there are still sufficient reasons to increase the ion source brightness and lower its energy spread. Apart from that, Xe$^+$ can also be generated from a single-atom tip GFIS, which shows high brightness with a more stable current and simplified design compared with the He GFIS\cite{lai2017xenon}.    
\par
New sources based on cold atoms have recently attracted attention in the research for accessing alternative ion species which aim at achieving higher brightness and smaller energy spread, including the magneto-optical ion source\cite{hanssen2006laser}, low-temperature ion source\cite{knuffman2013cold}, ColdFIB\cite{viteau2016ion}, and the atomic beam laser-cooled ion source (ABLIS)\cite{ten2017direct, ten2014performance}. For this type of source, neutral atoms undergo efficient sub-Doppler cooling by near-resonant lasers, and then are ionized by another focused laser to create a high brightness ion beam. Such technology can reduce energy spread and uses laser cooling to produce ion beams from over 27 elements\cite{mcclelland2016bright}, most of them not available in conventional FIB. Completed FIB systems with lithium\cite{gardner2019characterization}, chromium\cite{steele2011inter} and cesium\cite{viteau2016ion} have been constructed.
\par
Rb has also been studied for a long time as a potential ion species for FIB systems. Wouters \emph{et al} \cite{steinar2016, ten2017} successfully ionized Rb atoms with high brightness. The Rb FIB was then characterized to have imaging resolution of 3 nm and 0.2 eV energy spread at 8.5 keV, which is $10\times$ smaller than the energy spread of a Ga LMIS. Here, the investigation of the essential suitability of the ultracold Rb FIB for nanomachining applications is presented. First, simulations of ion-solid interaction are presented. Then, the prototype FIB system is introduced. After that, milling experiments are performed and milling rates of Rb$^+$ for various samples are measured. Finally, the results are summarized.

\section{\label{sec:exp}Modelling and experiments}
\subsection{Ion solid interactions}
In this section, the SRIM Monte Carlo software\cite{biersack1982stopping} is applied to understand the ion beam implantation and sputtering. Generally, the incident ions with primary energy collide with target atoms transferring momentum so that they leave their original lattice sites, causing damage to the target material. If the atom is near the surface, then it may escape from the target substrate, which is called sputtering. Sputter yield is used to indicate the number of target atoms sputtered by each incident ion. During the process, ions gradually lose energy and mostly stop inside the substrate. Ion range is used to describe the average distance of total incident ions from the substrate surface, while the lateral straggle characterizes the lateral distribution perpendicular to ion incidence.
\begin{figure}
    \includegraphics{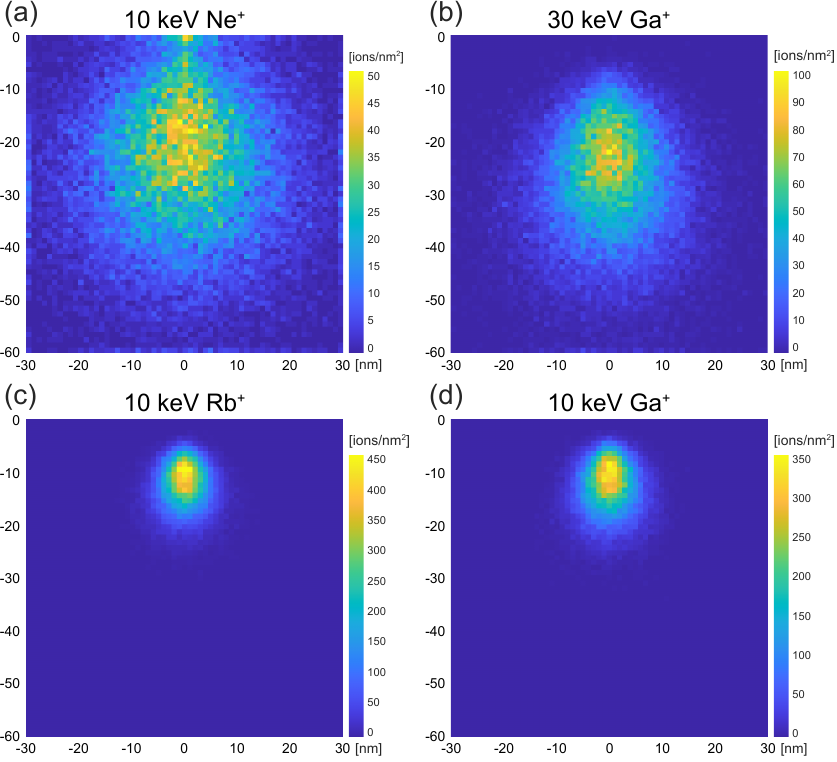}
    \caption{SRIM predictions for the ion solid interaction volume of three main ions into Si with perpendicular incidence. (a) 10 keV Ne$^+$; (b) 30 keV Ga$^+$; (c) 10 keV Rb$^+$; and (d) 10 keV Ga$^+$. The color bar shows the ion concentration.}
    \label{interaction volume}
\end{figure}

Figure \ref{interaction volume} presents the interaction volume of a few ion species into a silicon substrate as derived from SRIM. Here, three typical ions Ne$^+$, Ga$^+$, and Rb$^+$ are listed, and corresponding numerical results can be found in Table.\ref{table: SRIM simulation} of the Appendix. Fig. \ref{interaction volume}(b) shows the interaction volume of conventional 30 keV Ga$^+$, while (a), (c), and (d) are the three ion species of Ne$^+$, Ga$^+$, and Rb$^+$ with a beam energy of 10 keV. It can be observed that 30 keV Ga$^+$ has a large interaction volume with 27.0 nm ion range and 9.5 nm straggle, and that 10 keV Ne$^+$ has a similar volume with somewhat smaller ion range of 24.5 nm, but larger lateral straggle of 12.2 nm. This is mainly because of the low ion mass of Ne$^+$. For Ga$^+$ and Rb$^+$ at 10 keV, both show smaller interaction volumes, compared to 30 keV Ga$^+$. They share a similar ion range of \textasciitilde13 nm and small ion straggle. It can be seen that low beam energy reduces the range of ion damage to the substrate, but the sputter yield is also decreased according to the simulations.

\subsection{Ultracold FIB system}
\begin{figure}
    \centering
    \includegraphics{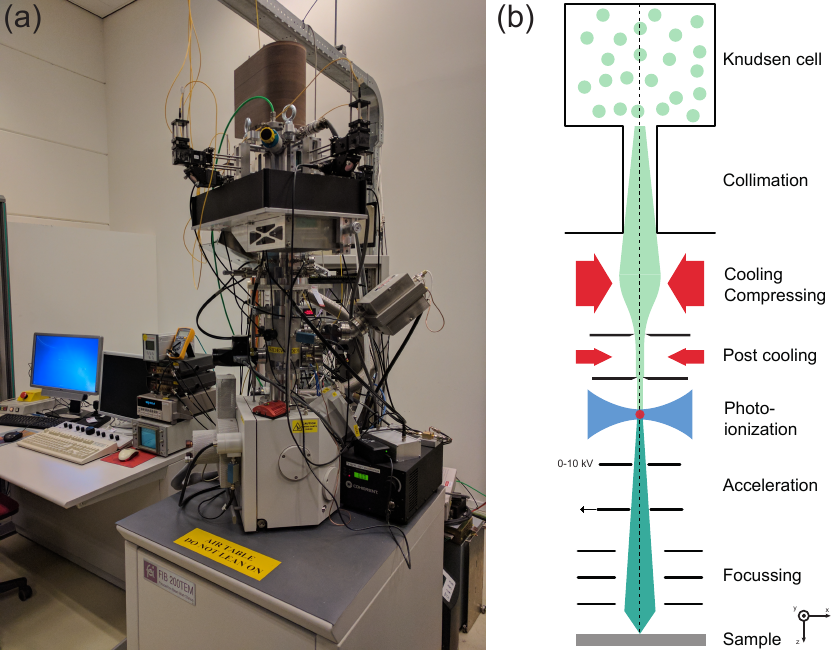}
    \caption{(a) An overview of Rb FIB in lab, consisting of ABLIS on top part mounted on the commercial FIB200TEM system. (b) Schematic of ultracold Rb FIB correspondingly.}
    \label{fig:NGFIB overview}
\end{figure}

Experiments are performed on a prototype ultracold Rb FIB system, as shown in Fig. \ref{fig:NGFIB overview}(a). 
For the instrument, the original Ga LMIS was removed and our Rb source is mounted instead onto the FEI FIB200TEM column with a custom-made connection assembly. External translators and supports are used to align the two columns and center the ion beam. Apart from the removal of the Ga LMIS, the whole FIB system is kept intact and maintains its complete functionality, including imaging and patterning.   

Fig. \ref{fig:NGFIB overview}(b) shows a schematic of the Rb system. Briefly, the Rb atoms effuse from the top Knudsen cell, and a long, thin tube works as a collimator to select atoms that can be laser cooled and compressed. Then these selective atoms enter a two-dimensional magneto-optical compressor (MOC). Inside the compressor, the atoms experience Doppler cooling and the temperature is decreased to around 200 $\mu K$ by two sets of circularly polarized laser beams while the beam is compressed in a magnetic field. Just below the MOC, an optical molasses is used for further cooling the transverse temperature of the Rb beam to near 40 $\mu K$, thus increasing the reduced brightness which is estimated to reach an equivalent value of $\rm 6\times 10^6  A/m^2\cdot sr\cdot eV$\cite{ten2017direct}. \par
Then the ultracold Rb atoms enter the accelerator equipped with an optical build-up cavity. The atoms can be excited by a focused excitation laser ($\lambda=780$ nm) and ionized by a cavity-enhanced ionization laser ($\lambda=480$ nm). An aperture, which is placed just above the ionization regime, is applied to select the Rb atoms to be ionized inside the cavity. After ionization, the ionized Rb$^+$ beam is accelerated by two electrodes placed above and below the cavity and sent into the conventional FIB ion column. A previous measurement has shown that for a 7 $\mu$m selective aperture, the final FIB images can achieve around 3 nm resolution at 0.2 eV energy spread\cite{ten2017}. \par
In this paper, a 50 $\mu m$ selective aperture is used which leads to an increased ion beam focus size of 160 nm ($d_{50}$) at the sample and an energy spread of 0.8 eV. It is noted that the ion-solid interaction mainly depends on the primary ion energy and ion mass. The Rb$^+$ ion beam is therefore suitable for studying the milling abilities despite the low resolution.

\begin{figure*}[ht]
    \centering
    \includegraphics{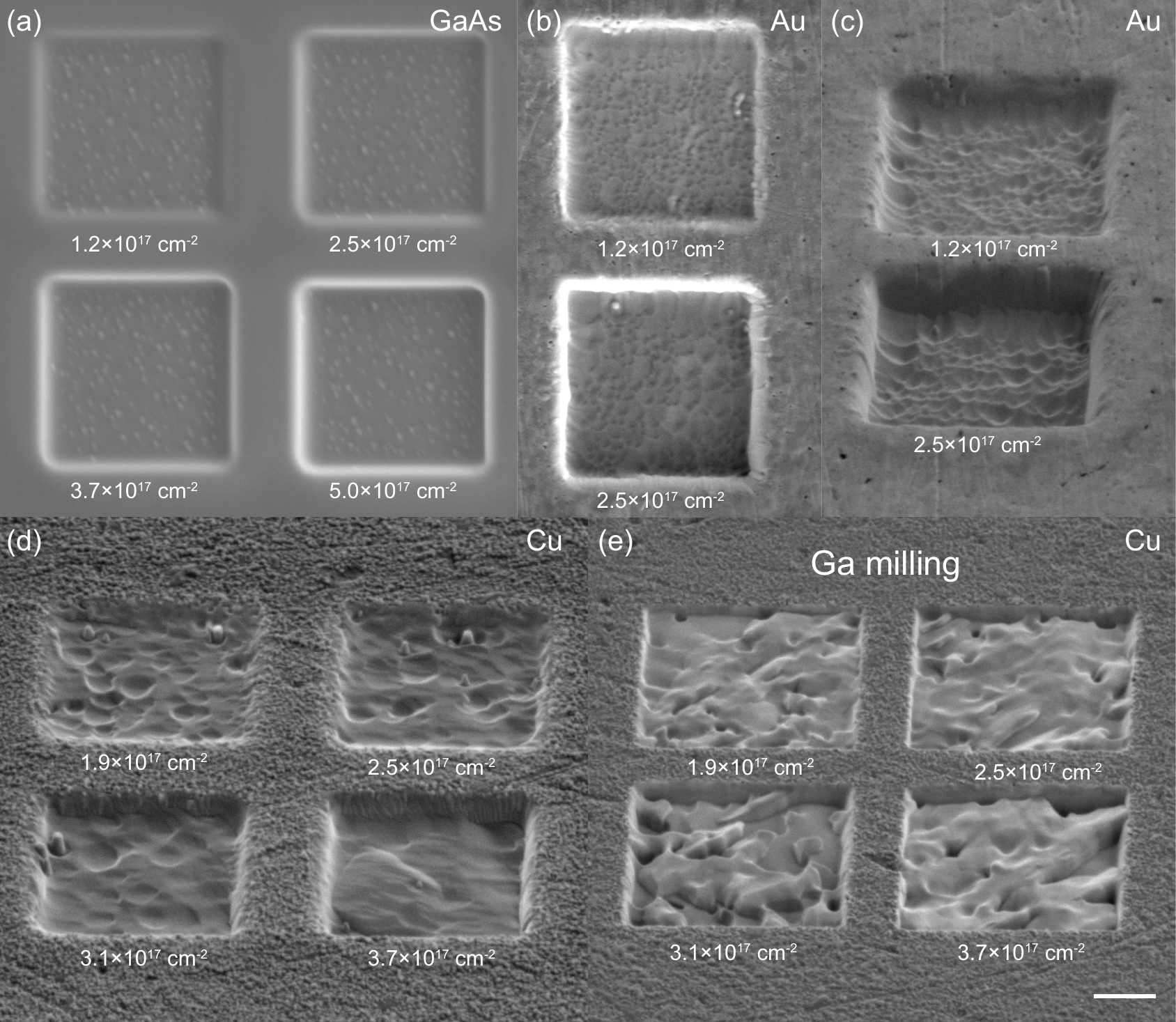}
    \caption{SEM images of Rb$^+$ and Ga$^+$ milled patterns (each pattern area of $3\times 3 \mu m^2$) on standard samples. (a) top view (0\textdegree) of milled patterns on GaAs; (b) top view (0\textdegree) of milled patterns on polycrystalline Au; (c) tilt view (52\textdegree) of (b); (d) tilt view (52\textdegree) of milled patterns on polycrystalline Cu; (e) tilt view (52\textdegree) of 30 keV Ga$^+$ milled patterns on polycrystalline Cu. Ion dose is marked below each pattern. Scale bar, 1 $\mu m$}
    \label{Milling patterns}
\end{figure*}

\subsection{Milling and characterization}
Milling experiments with 8.5 keV Rb$^+$ irradiation were performed in the FIB system described above with beam currents in the range 1-10 pA. Ga$^+$ milling was performed in a commercial FEI Nova Nanolab dual-beam FIB system equipped with a field-emission scanning electron microscope (SEM), which was used for imaging of the patterns created. The Ga$^+$ beam energy was set to 8.0 keV and 30 keV with beam current of 93 pA. Each milled pattern was repeatedly scanned with 50\% overlapping and 1 $\mu$s dwell time. All sample wafers used were obtained from Ted Pella Inc with ultrahigh purity and kept in high vacuum.

\section{\label{sec:result}Results and discussion}
In this section, milling experiments are performed under irradiation of 8.5 keV Rb$^+$, 8.0 keV Ga$^+$ and 30 keV Ga$^+$. Ion milling under 8.0 keV Ga$^+$ irradiation is done to compare with Rb$^+$ at the similar beam energy, while 30 keV Ga$^+$ serves as a reference.

\par
 Images of Rb$^+$ milled patterns for selected materials are shown in Fig. \ref{Milling patterns}. Here GaAs, Cu, and Au are used as standard sample materials and they were milled by the 8.5 keV Rb$^+$ ion beams with normal incident angle (0\textdegree) scanning. Fig. \ref{Milling patterns}(a) shows the result for GaAs of four different doses applied to four adjacent square patterns. The milling doses are $1.2\times10^{17}$, $2.5\times10^{17}$, $3.7\times10^{17}$, and $5.0\times10^{17}$ ions/cm$^2$. At the lowest dose of $1.2\times10^{17}$ ions/cm$^2$, nano droplets are clearly observed. With an increase in ion dose, there is limited change in droplet count and characteristic size. Xu \emph{et al}\cite{xu2016ion} reported similar nanodroplet findings for 10 keV Ga$^+$ milling on GaAs substrate, but the average size of nanodroplets is larger than that in this work. Milling patterns on polycrystalline Au are presented in Fig. \ref{Milling patterns}(b) and (c), top view (0\textdegree) and tilted view (52\textdegree) respectively. For the irradiation under different doses, the Au patterns show clear sidewalls after long milling times, and the milled surface has small fluctuations which can be confirmed in Fig. \ref{Milling patterns}(c). Fig. \ref{Milling patterns}(d) presents four adjacent patterns on polycrystalline Cu under increasing doses milled by Rb$^+$, while similar milling experiments were performed by 30 keV Ga$^+$ FIB as a comparison, shown in Fig. \ref{Milling patterns}(e). The doses of the four patterns are $1.9\times10^{17}$, $2.5\times10^{17}$, $3.1\times10^{17}$, and $3.7\times10^{17}$ ions/cm$^2$. In the first row in Fig. \ref{Milling patterns}(d), some craters exist on the patterned area. For the next higher dose of $3.1\times10^{17}$ ions/cm$^2$, the surface shows shallower craters at the bottom, and at the highest dose of $3.7\times10^{17}$ ions/cm$^2$, the surface tends to be smoother. In Fig. \ref{Milling patterns}(e) the four patterns milled by Ga$^+$ show larger roughness, especially apparent as a pitted landscape at the dose of $3.1\times10^{17}$ ions/cm$^2$. This phenomenon was observed and studied by other researchers of Ga$^+$ milling on polycrystalline Cu substrates\cite{casey2002copper}, which is caused by the channeling  effect\cite{kempshall2001ion}. One more experiment has also been done in the 8.0 keV Ga$^+$ FIB. The result shows that although milling with Ga$^+$ at lower beam energy has less surface roughness compared to 30 keV Ga$^+$, the surface still suffers from channeling effects when milled by Ga$^+$. From our observation on Rb$^+$ milling, such channeling is decreased.
\begin{figure}
    \centering
    \includegraphics{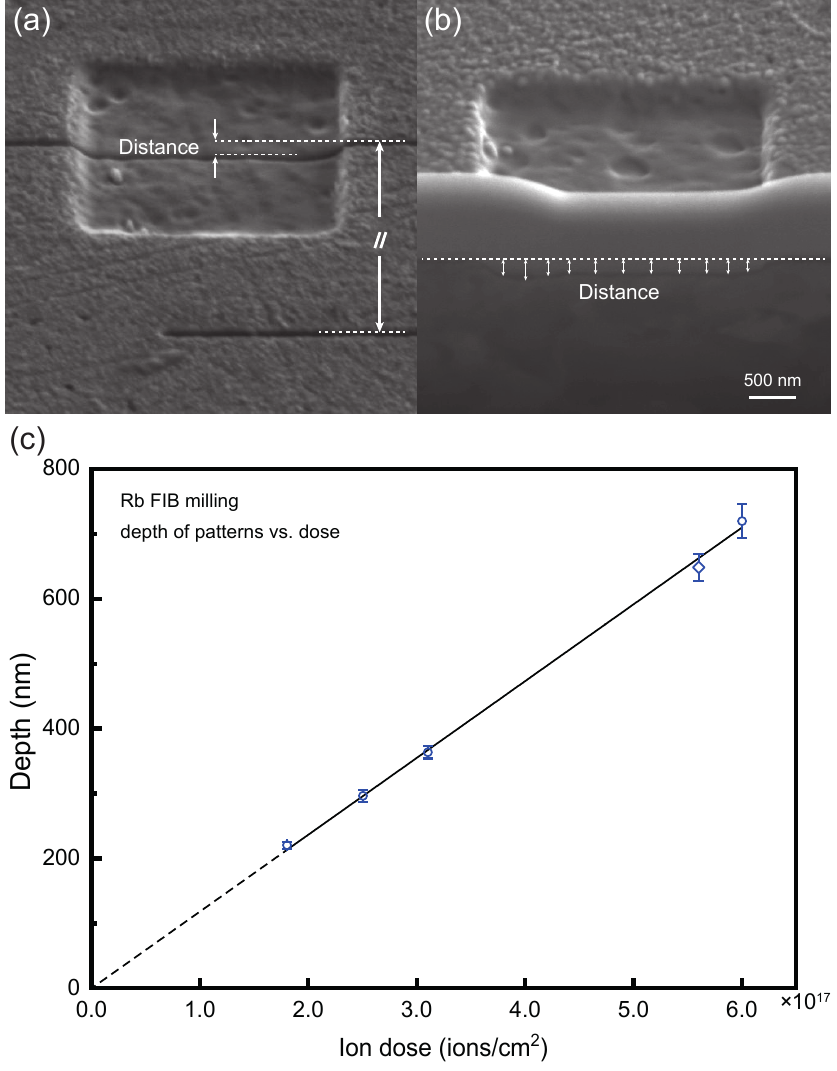}
    \caption{Methods to measure the depth of Cu patterns and calculate the sputter yield. (a) SEM image of Ga$^+$ line milling on Rb milled pattern (tilt view, 52\textdegree). Apply milling line profile across the pattern, and measure the distance between lines of surface and bottom.  The parallel line works as reference. (b) SEM image of cross section of Rb$^+$ milled pattern (tilt view, 52\textdegree). Firstly deposit platinum on the pattern as protection layer, and then make cross section to measure the distance. (c) Plots of experimental data and their corresponding linear fitting function. Points with circles represent data acquired from line-burn method, while point with diamond marker represent data acquired from cross-section method.}
    \label{Cu sputter}
\end{figure}
\par
 To determine the sputter rates of Rb$^+$ on each substrate, two methods are applied to measure the pattern depth. Figure \ref{Cu sputter} shows the depth measurement and sputter rate calculation result on a copper substrate as an example. The first method is the line-burn method, shown in Fig. \ref{Cu sputter}(a). Two parallel lines are milled on the substrate using the Ga FIB with one line near the target pattern and the other one across it. Then the sample is observed at 52\textdegree\ by SEM and the distances between the lines are measured. In most cases, the line-burn measurement is applied due to its quick test. The second method involves making a cross section in the middle of the milling pattern, shown in Fig. \ref{Cu sputter}(b). Electron- and ion-induced platinum deposition are applied to protect the pattern surface. The cross section offers a direct view to measure the perpendicular distance from the substrate surface to pattern features. For both methods, more than 10 absolute distances between pattern features and substrate surface are acquired, and the averaged distance is then taken as pattern depth. Fig. \ref{Cu sputter}(c) shows measured depth versus ion dose for the sputter rate analysis of Rb$^+$ milling on Cu as an example. Five patterns on a copper substrate are milled under different ion doses. Four are measured by the line-burn method, shown as circles, while one is measured by the cross-section method, shown as a diamond marker. The data points range from low ion dose to high ion dose so that the slope can be determined. The linear fit shows the five data points match well, and the error bar for each data point represents the standard error of the depth measurements for each pattern. The fitted slope is $0.73\pm 0.06$\ $\mu$m$^3$/nC, where 0.73 is the sputter rate of Rb$^+$ on Cu, and 0.06 is the standard error. The margin of error is 0.04 with 95\% confidence level. This uncertainty of 5.9\% is typical for all reported sputter rates.

\begin{table*}
\centering
\caption{Sputter rate measurements for Rb$^+$ ions at 8.5 keV and Ga$^+$ ions at 8 keV, accompanied with simulation result for Rb$^+$ ions at 8.5 keV from SRIM software and reference sputter yields from other works}
\begin{ruledtabular}
\begin{tabular}{cccccccc}
\multicolumn{1}{c}{Material} &
\multicolumn{3}{c}{Measurements ($\mu$m$^3$/nC)}  &
\multicolumn{3}{c}{Simulations($\mu$m$^3$/nC)} &
\multicolumn{1}{c}{Other work} \\
\cline{2-4} \cline{5-7} \cline{8-8}
  & \multicolumn{1}{c}{Rb$^+$ 8.5 keV}  &  \multicolumn{1}{c}{Ga$^+$ 8.0 keV} &  \multicolumn{1}{c}{Ga$^+$ 30 keV} 
 & \multicolumn{1}{c}{Rb$^+$ 8.5 keV} &  \multicolumn{1}{c}{Ga$^+$ 8.0 keV} &  \multicolumn{1}{c}{Ga$^+$ 30 keV}
  & \multicolumn{1}{c}{Ga$^+$ 30 ke$^+$V}\\
\hline
B & 0.09 & 0.05 & 0.07 & 0.07 & 0.06 & 0.09 & \\
Diamond & 0.41 & 0.08 & 0.16 & 0.06 & 0.05 & 0.07 & \begin{tabular}[c]{@{}c@{}}0.09\cite{burnett2016large}\\ 0.18\cite{orloff1996fundamental}\end{tabular} \\
Al & 0.11 & 0.28 & 0.32 & 0.33 & 0.31 & 0.40 & \begin{tabular}[c]{@{}c@{}}0.31\cite{burnett2016large}\\ 0.28\cite{ostadi2009characterisation}\end{tabular} \\
Si & 0.22 & 0.17 & 0.27 & 0.21 & 0.20 & 0.22 &  \begin{tabular}[c]{@{}c@{}}0.27\cite{orloff1996fundamental}\\ 0.25\cite{ostadi2009characterisation}\end{tabular}\\
Ti & 0.21 & 0.12 & 0.32 & 0.20 & 0.19 & 0.25  & 0.31\cite{burnett2016large} \\
Cu & 0.73 & 0.41  & 0.60 & 0.60 & 0.60 & 0.74 & 0.15-0.55\cite{burnett2016large} \\
Au & 1.39 & 1.30 & 1.50 & 1.62 & 1.39 & 1.84 & 1.50\cite{orloff1996fundamental} \\
GaAs & 0.81 & 0.59 & 0.80 & 2.24 & 2.25 & 2.53 & \begin{tabular}[c]{@{}c@{}}0.86\cite{burnett2016large}\\ 0.61\cite{orloff1996fundamental}\end{tabular}   \\
InP & 1.45 & 0.83 & 1.40 & 1.67 & 1.56 & 1.87 &   \\
\end{tabular}
\label{table: Sputter yield table}
\end{ruledtabular}
\end{table*}
 
 %\begin{figure}
 %   \centering
 %   \includegraphics{Sputter yield measurements(3.375 inches width).pdf}
%    \caption{Results of sputter yield measurement of Rb and Ga milling. Red spot represent sputter yields of 8.5 kV Rb, blue diamonds shows the results of 8.0 kV Ga, and black triangles are the sputter yields of conventional Ga at 30 kV. The error bars of Rb are acquired by the linear fitting function of depth and ion doses for each substrate. }
%    \label{sputter yield}
%\end{figure}

By extracting the depth from measurements, the sputter rate for each substrate can be determined by dividing by the corresponding ion dose. The result is shown in Table.\ref{table: Sputter yield table}. The experimental results are presented in the first main column, with three sub-columns of 8.5 keV Rb$^+$, 8.0 keV Ga$^+$, and 30 keV Ga$^+$, as the main purpose is to compare the Rb$^+$ milling performance with the standard Ga LMIS. The second main column consists of the corresponding simulation results, which serve as a prediction of sputtering rates. And the last column is sputter rates of 30 keV Ga$^+$ from the literature. The consistency between our measurements of 30 keV Ga$^+$ and the reference results gives confidence that the substrates and measurements are trustworthy. 
\par
The experimental results between 8.5 keV Rb$^+$ and 8.0 keV Ga$^+$ are firstly compared. It is clear that, in most cases, Rb$^+$ shows higher sputter rates under similar beam energy, except for Al. Since sputter rate is mainly dependent on the ion species, beam energy, incident angle, and the target material\cite{nastasi1996ion}, Rb$^+$, with heavier ion mass, should have higher sputter rates. This is also revealed by the simulation results. When it comes to the comparison between 8.5 keV Rb$^+$ and 30 keV Ga$^+$, the results differ in various materials. For Al, Si, Ti, and Au substrates, the sputter rates of our Rb$^+$ are lower than that of conventional Ga$^+$, which is consistent with simulation results. In contrast, on B, diamond, Cu, GaAs, and InP, the Rb$^+$ beam shows larger material removal rates than Ga$^+$.
\par 
There are several noteworthy results to point out. One remarkable result is the Rb$^+$ sputter rate on the diamond substrate, 0.41 $\mu$m$^3$/nC, around 2.6 times larger than that for 30 keV Ga$^+$, while the simulation result is quite low, 0.06 $\mu$m$^3$/nC. One more controlled experiment was therefore performed on a conducting carbon substrate. The sputtering rate of 8.5 keV Rb$^+$ on this carbon substrate is 0.31 $\mu$m$^3$/nC, which is also beyond expectation. Both measurements indicate that Rb$^+$ beams have a more powerful material removal ability on carbon-based material. Al, as an outlier in the result table, in contrast shows a sputter rate of 0.11 $\mu$m$^3$/nC for 8.5 keV Rb, $2\times$ times lower than that of Ga for similar beam energy and $3\times$ times lower than that for normal 30 keV Ga. This result does not meet the expectation of SRIM and indicates that the Rb ion beam has poor milling performance on aluminum or oxidized aluminum. But then this could also be an advantage for ion microscopy on specific aluminum inter-connections or substrates with an aluminum protection layer. The milling behavior on aluminum may require further study.
\par
When focusing on GaAs, one can find for both Rb$^+$ and Ga$^+$, that there are discrepancies between measurements and simulations. This is mainly due to the default model and surface binding energy used in the SRIM software. Seah \emph{et al}\cite{seah2010sputtering} has pointed out that a semi-empirical model would be more practical to find the right parameters for the SRIM simulation on GaAs. By using the updated parameters from their research, the simulation result of 8.5 keV Rb$^+$ is 1.12 $\mu$m$^3$/nC, which is closer to the experimental result, 0.81 $\mu$m$^3$/nC.
\par
In summary, in the majority of standard substrates used in experiments, Rb$^+$ shows good sputter yield ability, even compared with 30 keV Ga$^+$. It should be noted that the Rb FIB system uses low beam energy, which has the advantage of a short projection range of Rb$^+$ into the substrates. For example, the sputter rates for our Rb$^+$ and 30 keV Ga$^+$ on Si are quite similar (8.5 keV Rb$^+$: 0.22 $\mu$m$^3$/nC, and 30 keV Ga$^+$: 0.27 $\mu$m$^3$/nC), but Rb$^+$ has only around half the ion range of 30 keV Ga$^+$ (see Table.\ref{table: SRIM simulation}). It has been confirmed that heavier ions can lead to low ion implantation range, such as Xe\cite{burnett2016large}. Further study will be focused on the Rb$^+$ implantation.

\section{Conclusions}
A prototype FIB system has been applied to investigate the potential milling application of ultracold Rb ions. The measurements for various standard metal and semiconductor materials indicate that Rb$^+$ ions in general have somewhat higher sputter rates compared with Ga$^+$ ions at the same energy. It is also found that a Rb$^+$ ion beam has a substantially quicker carbon substrate removal rate but a slower milling rate on aluminum, which may meet specific demands for the two substrates. Overall this investigation indicates that Rb is a realistic alternative to Ga in milling applications for typical materials when a sufficient and comparable milling rate is the main concern.

\begin{acknowledgments}
This work is part of the project Next-Generation Focused Ion Beam (NWO-TTW16178) of the research programme Applied and Engineering Sciences (TTW) which is (partly) financed by the Dutch Research Council (NWO). The authors would like to acknowledge helpful discussions with Greg Schwind, Chad Rue, Erik Kieft and Yuval Greenzweig.
\end{acknowledgments}

\section*{Author Declarations}
\subsection*{Conflict of Interest}
The authors have no conflicts to disclose.

\section*{Data Availability Statement}

The data that support the findings of this study are available from the corresponding author upon reasonable request.

%\clearpage

\appendix

\section{SRIM simulation}
The simulation is completed with the SRIM-2013 software. The damage type is chosen as "Detailed Calculation with full Damage Cascade". For each calculation, the total number of ions is set to 35000, and layer thickness is set to 80 nm. The resulting values of sputter yield, ion range and lateral straggle for Si substrates are collected in Table. \ref{table: SRIM simulation}.

\begin{table}[h]
\centering
\caption{SRIM simulation results of Ne$^+$, Ga$^+$, and Rb$^+$ into silicon substrate. Total number of incident ions simulated is 35000 in each case.}
\begin{ruledtabular}
\begin{tabular}{ccccc}
    Ion species &  \begin{tabular}[c]{@{}c@{}}Beam energy \\ (keV) \end{tabular} & \begin{tabular}[c]{@{}c@{}}Sputter yield \\ (atoms/ion)\end{tabular} & \begin{tabular}[c]{@{}c@{}}Ion range \\ (nm)\end{tabular} & \begin{tabular}[c]{@{}c@{}}Straggle \\ (nm)\end{tabular} \\
    \hline
    Ne$^+$ & 10 & 1.16 & 24.5 & 12.2 \\
    Ga$^+$ & 30 & 2.29 & 27.0 & 9.5 \\
    Ga$^+$ & 10 & 1.69 & 13.5 & 5.0 \\
    Rb$^+$ & 10 & 1.77 & 12.8 & 4.3 \\
\end{tabular}
\label{table: SRIM simulation}
\end{ruledtabular}
\end{table}
\par

\nocite{*}
\bibliography{ref}% Produces the bibliography via BibTeX.

\end{document}